# Nitrogen-induced Local Spin Polarization in Graphene on Cobalt

Zhongping Chen[1*], Ling Miao[1], Xiangshui Miao[1,2†]

[1] *Department of Electronic Science and Technology, Huazhong University of Science and Technology, 430074 Wuhan, China*

[2] *Wuhan National Laboratory for Optoelectronics, 430074 Wuhan, China*

By means of first principles calculations we demonstrate an effective method to tailor the local spin configuration of graphene on Co(0001) surface through nitrogen doping. Two different site occupancies of the N impurities are discussed with the focus on structural, electronic and magnetic properties. N induces opposite local spin polarization at the two sites through $\pi$-$d$ Zener exchange-type hybridization with Co substrate. In addition, the induced spin polarization is energy dependent and controllable by electric field. Consequently this structure can be applied as a spin injection source in graphene based spintronics.

## I. INTRODUCTION

Since its first experimental discovery,[1] graphene has become of considerable interest to scientists in many areas due to its intriguing physical properties.[2,3] Bing comprised of light element C with weak spin-orbit coupling, graphene enjoys a large spin relaxation length (about 2 $\mu$m) at room temperature,[4] which makes it an exceptional spin transport medium and advantageous over conventional semiconductors in spin electronics (spintronics) application.[5] One crucial issue for the achievement of graphene based spintronics is how spin-polarized electrons can be effectively injected into nonmagnetic graphene. Ferromagnetic (FM) contact is among the most popular spin injection methods applied in graphene spin valve devices.[6-8] Combining graphene and FM metals, such as Co and Ni, has been studied with the focus on interfacial structural and electronic properties,[9-14] and predicted to have promising applications. Nevertheless, experimental results have shown that magnetoresistance (MR) ratios in such spin valves are fairly low.[6,]

[15] Namely the spin injection efficiency at the graphene/FM-metal interfaces is rather limited. Therefore, it is of particular significance to find other efficient ways to inject spin into graphene.

In this report, we theoretically propose one highly efficient spin injection method by implanting N impurity into a graphene sheet that is grown on Co(0001) surface. So far, there have been several experimental approaches for doping nitrogen in graphene, including chemical vapor deposition,[16, 17] ion implantation[18] and plasma processing[19]. Simultaneously, nitrogen impurities in graphene have been studied theoretically as well.[20-23] These results have shown that nitrogen in graphene layers are almost nonmagnetic namely not spin-polarized,[20, 22] except some edge doping situations.[22, 23] Here, by employing Co substrate, we are able to turn N impurity to highly spin-polarized states and thus create highly efficient spin injection sources in graphene. Furthermore, the spin polarization induced by N is localized, so that it can contribute to point spin injection source in nanospintronics and spin quantum bit in quantum computation. Hereby, our method is a prototype and will inspire new approaches for tailoring the local spin configuration of graphene through point defects with the help of magnetic substrates.

## II. COMPUTATIONAL DETAILS

Our theoretical study is based on first principles calculations at the level of spin-polarized density functional theory (DFT),[24, 25] using projector augmented wave (PAW) formalism[26] as implemented in VASP code[27]. General gradient approximation (GGA) with Perdew–Burke–Ernzerhof (PBE) exchange-correlation function[28] is employed together with a plane wave cutoff of 400 eV. Brillouin zone integration is performed on a Γ centered $6\times6\times1$ grid using Methfessel-Paxton scheme.[29] The structural models are constructed by substituting one C of a $3\times3$ graphene supercell with one N and covering the graphene sheet over a Co(0001) substrate of five Co layers. The in-plane lattice constant $a$ is fixed to 2.506 Å, the same as Co(0001) lattice, and the vacuum slabs are set to more than 10 Å above the graphene sheets. Atoms in the graphene sheet as well as the upper most two Co layers are free to move during atomic relaxation, which is conducted by conjugate gradient method,[30] and the structures are optimized when the force acting on each atom is less than 0.01 eV/Å.

**III. RESULTS AND DISCUSSION**

Firstly the characteristics of pure graphene adsorbed on Co surface are briefly reviewed.[31] The lattice mismatch between graphene and Co(0001) plane is so small (1.8%),[32] that the graphene sheet is in registry with the Co surface and suffers only a very little tensile. Above the Co(0001) surface, there are three sites marked *A*, *B* and *C* in Fig. 1(a). Two of them may be occupied by Graphene C atoms, resulting in three on-top registry configurations named $Gr^{AB}$, $Gr^{AC}$ and $Gr^{BC}$. Our results show that $Gr^{AC}$ is the most stable one with graphene-Co separation $d_0$ = 2.12 Å. The projected density of states (PDOS) of C atoms in $Gr^{AC}$ is presented in Fig. 1(b). All s, $p_x$ and $p_y$ components are symmetric and only $p_z$ component is asymmetric with respect to the two spins. This indicates $\pi$-*d* hybridization between out-of-plane $\pi$ ($p_z$) states of graphene and *d* band of Co substrate. Because of their different site occupancies, the two C atoms present different $p_z$ PDOS features. $C^A$, sitting above a surface Co, presents positive spin peaks within [−2.8, −2.2] eV, [0.1, 1.2] eV intervals, and negative spin peaks within [−1.7, −1.0] eV, [1.4, 2.2] eV intervals. In comparison, $C^C$ presents a strong positive peak within [−1.0, −0.2] eV, a strong negative peak and a week positive peak within [0.6, 1.4] eV. It is noteworthy that PDOS of both $C^A$ and $C^C$ diminish at the Fermi level ($E_F$). Therefore the graphene is not spin-polarized[33] at $E_F$, and the resistance across the interface is large. This is probably a reason for the low spin injection efficiency at the interfaces in the spin valves.

Next, we study the effects of N impurity in $Gr^{AC}$. By substituting one $C^A$ or $C^C$ with N, we create two situations $Gr^{AC}N^A$ and $Gr^{AC}N^C$ as shown in Fig. 2(a) and (b) respectively. Some data determining their properties is presented in Table I. In either case N is located above the graphene layer with the distance $d_1$ larger than $d_0$. N is more likely to substitute $C^C$ since the formation energy is a little higher.

PDOS of $N^A$ impurity in $Gr^{AC}N^A$ is presented in Fig. 2(g). The out-of-plane $p_z$ component becomes spin asymmetric above $-5$ eV. For instance, there is only a positive PDOS peak at $E_F$, while in the negative spin channel the PDOS is almost zero at $E_F$. A sharp positive peak and a sharp negative peak are also presented at $-2.8$ eV and $-1.8$ eV respectively. Similarly, $N^C$ displays spin asymmetric $p_z$ PDOS as well [see Fig. 2(h)], but the characteristics are completely different from $N^A$. In the vicinity of $E_F$, the $N^A$ positive peak vanishes in $N^C$ PDOS. Instead, a broad negative peak appears within $[-0.6, 0.1]$ eV. A strong positive peak is presented within $[-1.6, -0.8]$ eV and two other positive peaks are located around 0.3 eV and 1.6 eV. The $\pi$-$d$ Zener exchange mechanism[34] can be employed to explain the spin asymmetry of $p_z$ states of these N impurities. In diluted magnetic semiconductors, the $p$ band of semiconductor is mixed with the $d$ states of the transition metal impurity ions and thereby broadened and shifted with different variations in the two spin channels around $E_F$. In the cases discussed here, Co substrates play the roles of transition metal impurity and N act as the semiconductor. Specifically, in $Gr^{AC}N^A$, strong hybridization is apparent as exhibited in Fig. 2(g): the positive peak at $-2.8$ eV and the negative peak at 1.2 eV are present both in the $p_z$ PDOS of on-top $N^A$ and the $d_{xz}$, $d_{yz}$ and $d_{z^2}$ PDOS of surface $Co^A$. While in $Gr^{AC}N^C$, the $N^C$ negative peak around $E_F$ is the result of strong $\pi$-$d$ hybridization with $d_{xz}$ and $d_{yz}$ states of the nearest Co ($Co^A$) [See Fig. 2(h)]. The positive peak, as well as the negative states, within $[-1.6, -0.8]$ eV also hybridize with Co $d_{xz}$ and $d_{yz}$ states in the same energy interval. Below $-5.5$ eV in Fig. 2(g) and (h), where Co PDOS drops to zero, both $N^A$ and $N^C$ display spin symmetric $p_z$ PDOS because of the absence of $\pi$-$d$ mixing.

It may be assumed that the N impurity will introduce some changes to its neighboring graphene C and substrate Co. In fact, this is true only when N takes the place of on-site $C^A$ in $Gr^{AC}N^A$. $N^A$ hybridizes substantially with its neighboring C in the conjugate loops and causes considerable changes to their PDOS [see Fig. 2(g)]. Most apparently, all $C_1^C$, $C_2^A$ and $C_3^C$ present the strong positive peaks at $E_F$ as $N^A$ does, which is not found in undoped $Gr^{AC}$. Also the positive peak around $-2.8$ eV and the negative peak around $-1.8$ eV exist in $N^A$ PDOS due to the hybridization between $N^A$ and $C_2^A$ that are both on-top of $Co^A$. $N^A$ brings some changes to the $Co^A$ beneath it too. Especially for the $d_{z^2}$ PDOS, in Fig. 2(g) the positive peak within $[-1.5, -0.4]$ eV and the negative peak within $[0.2, 1.5]$ eV are both broadened and lifted compared to the $Co^A$ in $Gr^{AC}N^C$ as shown in Fig. 2(h). Nevertheless, in the more stable $Gr^{AC}N^C$, the PDOS characteristics of neighboring C and Co are hardly altered. For example, the first (third) nearest neighbor $C_1^A$ ($C_3^A$) of $N^C$ is characterized by two positive PDOS peaks within $[-2.8, -2.2]$ eV, $[0.1, 1.2]$ eV and two negative peaks within $[-1.8, -1.0]$ eV, $[1.3, 2.2]$ eV [see Fig. 2(h)], which is the hallmark of $C^A$ in pure $Gr^{AC}$ on Co. The second nearest neighbor $C_2^C$ presents similar PDOS as the $C^C$ shown in Fig. 1 too. The $\pi$-$d$ hybridization features of undoped $Gr^{AC}$ with Co substrate (not shown here) are also retained, such as the positive (negative) hybridization between $Co^A$ and $C_1^A$ within $[-2.8, -2.2]$ eV ($[-1.8, -1.0]$ eV), and the positive hybridization between $Co^A$ and $C_2^C$ around $-0.5$ eV [see Fig. 2(h)].

Under the influence of Co substrate, N impurity introduces high spin polarization[33] around $E_F$ to the graphene layer and thus substantially increases the initially diminishing density of states around $E_F$ in one spin channel. [Note that the vertical scales of N PDOS are twice the scale of C in Fig. 2(g) and (h)] So that the electrons carrying one certain spin can easily travel into the graphene but those carrying the other spin are still blocked at the interface. This mechanism exactly exhibits the function of an effective spin injection source. The spin polarization induced by N impurity is localized since the PDOS of those C far away from N are found to be unaffected. Hence, this spin injection source can be viewed as a dimensionless point compared to the graphene sheet, which possibly contributes to graphene based nanospintronics. Relatively speaking, the spin-polarized area in $Gr^{AC}N^A$ is larger than in $Gr^{AC}N^C$, since not only $N^A$ but also all the neighboring C atoms in the three conjugate loops are positively spin-polarized at $E_F$, and accordingly the spin-polarized current in $Gr^{AC}N^A$ is expected to be higher. However, the spin polarization at $E_F$ of $Gr^{AC}N^A$ is inverted with respect to the Co substrate, while $Gr^{AC}N^C$ has the same spin polarization at $E_F$ as Co [see Fig. 2(g), (h)]. Therefore which configuration is more favorable for a spin injection source still needs to be determined experimentally. Although $Gr^{AC}N^C$ is more stable according to Table I, the formation energy difference here (74.82 meV) is rather tiny when compared to those appearing in the literature (~ 1 eV).[20, 23] Also, scanning tunneling microscopy (STM) study of graphene on Co has shown that the *A* and *C* sites are distinguishable.[9] Therefore it is possible to artificially control the implantation of N impurity into either site by STM, and to create the desirable spin injection source.

It is intriguing that the spin polarization of N impurity is energy-dependent. Within a certain energy interval where a positive or a negative peak presents, the N states are accordingly positively or negatively spin-polarized [see Fig. 2(g) and (h)]. The spin polarization in each of these energy intervals is quite high and sometimes approaches 100%. This feature makes it possible to control the polarization direction of the spin injection source by electric field, in addition to conventional magnetic control. When an electric field is applied perpendicular to the surface by a gate voltage, $E_F$ is shifted proportionally to the voltage. Therefore the spin polarization can be tuned simply by shifting $E_F$ into a certain energy interval of the desired spin polarization.

To further discuss the effects of different N site occupancies, we present in Fig.2 the spin density and charge density in certain planes of the two $Gr^{AC}N^A$ and $Gr^{AC}N^C$ supercells. $N^A$ presents a diminishing positive spin density in Fig. 2(c), so its magnetic moment is almost zero. The three nearest $C^C$ keep the positive spin density and thus yield a positive net magnetic moment in the vicinity of $N^A$. While, $N^C$ presents a considerable negative spin density in Fig. 2(d) with a magnetic moment of $-0.04$ $\mu_B$ (see Table I) and gives rise to a negative net magnetic moment together with its three $C^A$ neighbors. Since the electronegativity of N (3.04) is larger than C (2.55), $N^A$ attracts more charges from $Co^A$ than $C^A$ does. So when N is displaced from hollow site $C$ to on-top site $A$, $e_N$ increases while $e_{Co}$ decreases, causing corresponding changes to the magnetic moments as presented in Table I. Accordingly the bonding between $N^A$ and $Co^A$ is more ionic than C-Co bonding, as seen from Fig. 2(e) that the overlap of charge density between $N^A$ and $Co^A$ is slighter than that between $C^A$ and the underlying $Co^A$. The interaction between $N^A$ and $Co^A$ is a little repulsive so that $N^A$ is located slightly above the graphene layer and $Co^A$ is slightly pushed downwards [see Fig. 2(e)], which causes the N-Co distance $d_2$ to be a little larger than $d_1$ in Table I. In comparison, $N^C$ is located above the hollow site surrounded by three $Co^A$ with a larger N-Co distance ($d_2$ = 2.85 Å) in $Gr^{AC}N^C$. So the interaction between $N^C$ and $Co^A$ is weaker and indirect as shown in the charge density in Fig. 2(f) that the overlap is very slight and, to some degree, in the in-plane direction. These findings are consistent with the PDOS properties discussed above. For instance, $N^A$ hybridizes strongly with all out of plane $d$ states, especially $d_{z^2}$ states, of $Co^A$. While $N^C$ only hybridizes with $Co^A$ $d_{xz}$ and $d_{yz}$ states, leaving $d_{z^2}$ states unchanged.

We have also examined N impurity in $Gr^{AB}$ and $Gr^{BC}$. In $Gr^{AB}$, $N^A$ and $N^B$ behave just similarly as $N^A$ and $N^C$ in $Gr^{AC}$ respectively. While in $Gr^{BC}$, both $N^B$ and $N^C$ show the identical symmetric PDOS regarding the two spins, and their magnetic moments, as well as moments of all the C atoms, are zero. This is because the hybridization effect disappeared with a large graphene-Co separation ($d_0$ = 4.18 Å). Therefore, the influence from the Co substrate is indispensable in generating the spin polarization in N doped graphene.

## IV. CONCLUSION

In summary, we have theoretically demonstrated that the local spin configuration of graphene sheet on Co(0001) surface can be tailored by N substitution. N is positively spin-polarized at *A* site while negatively spin-polarized at *C* site. The spin polarization in either case is as high as almost 100% and largely energy dependent. Therefore they act as spin injection sources that can be controlled electrically as well as magnetically. *A* site substitution brings more modification to the PDOS of local atoms than *C* site substitution and contributes to the positive spin polarization of the surrounding C. N impurity introduces positive and negative net magnetic moments to the occupied *A* site and *C* site respectively, which can be regarded as spin quantum bits as well.

## ACKNOWLEDGEMENTS

This work is funded by National Natural Science Foundation of China (No. 50871043). The computational resources are provided by Wuhan National Laboratory for Optoelectronics (WNLO).

[*]Present address: Department of Physics, The University of Texas at Austin, Austin, Texas 78712, USA

zpchen@physics.utexas.edu

[†]miaoxs@mail.hust.edu.cn

TABLE I. Comparison of formation energy relative to $Gr^{AC}N^{C}$ [$\Delta E_f$ (meV)], height of N above Co surface $d_1$ (Å), the nearest N-Co distance $d_2$ (Å), charge population (e) and magnetic moment ($\mu_B$) of N ($e_N$, $\mu_N$) as well as neighboring Co ($e_{Co}$, $\mu_{Co}$) in two N substitution situations $Gr^{AC}N^{A}$ and $Gr^{AC}N^{C}$ (superscripts represent site occupancies).

|  | $\Delta E_f$ | $d_1$ | $d_2$ | $e_N$ | $\mu_N$ | $e_{Co}$ | $\mu_{Co}$ |
|---|---|---|---|---|---|---|---|
| $Gr^{AC}N^{A}$ | 74.82 | 2.25 | 2.39 | 3.16 | 0.00 | 7.49 | 1.80 |
| $Gr^{AC}N^{C}$ | 0.00 | 2.42 | 2.85 | 3.15 | −0.04 | 7.50 | 1.52 |

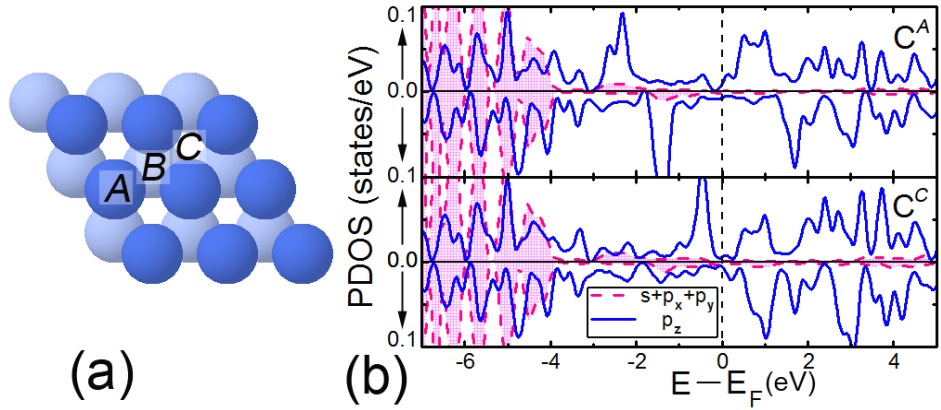

FIG. 1 (color online) Pure graphene on Co. (a) Top view of $3\times3$ Co surface supercell with three sites marked *A*, *B* and *C*. Dark blue spheres represent surface Co atoms ($Co^{A}$), and light blue ones represent the Co atoms in the second monolayer ($Co^{B}$). (b) Spin-resolved projected density of states (PDOS) of the C atoms in $Gr^{AC}$ configuration. The superscript of the notation in each pane denotes the site occupancy of the C atom.

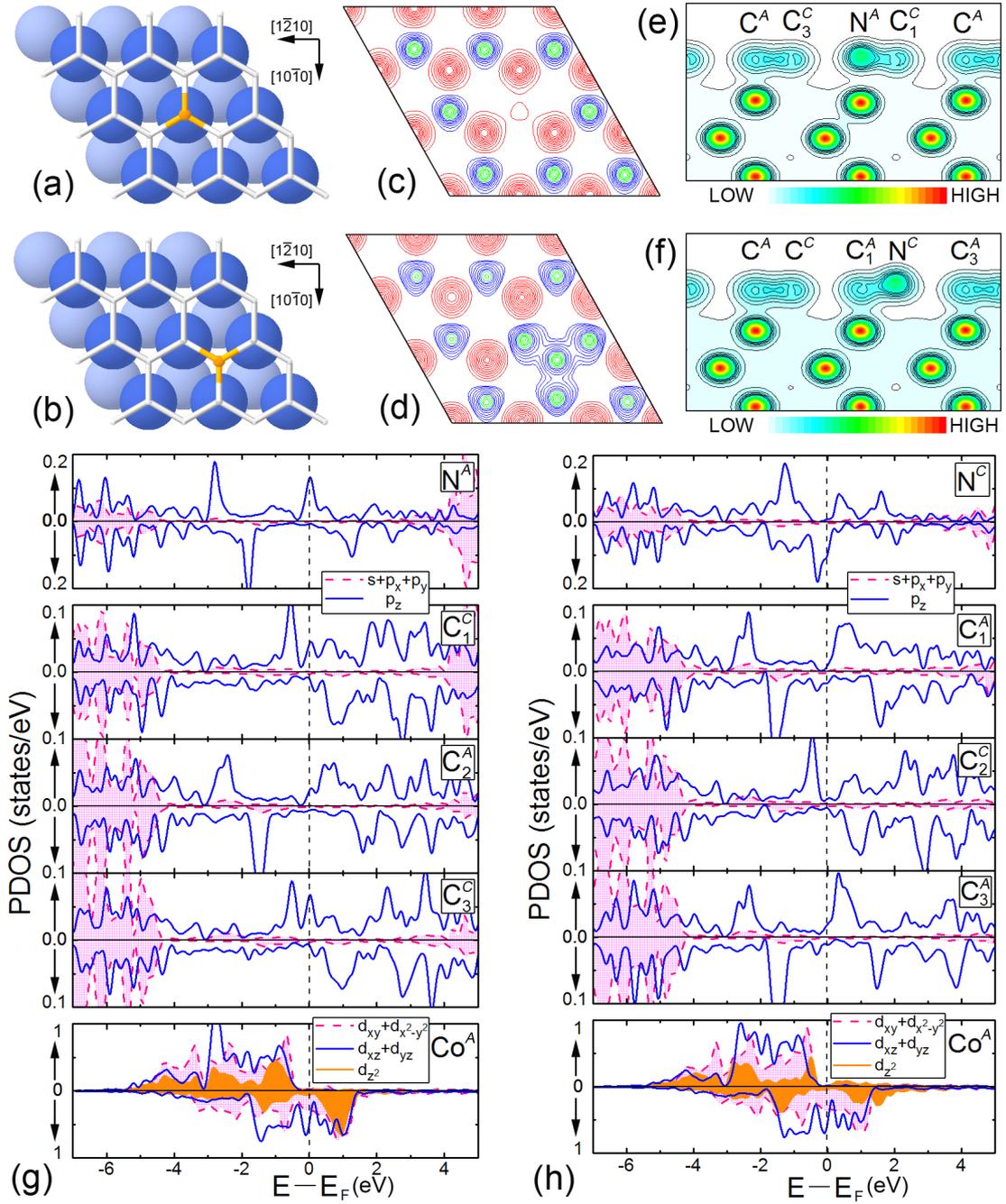

FIG. 2 (color online). In-plane structural model of (a) $Gr^{AC}N^A$ and (b) $Gr^{AC}N^C$. The gray grids represent the graphene lattice and the orange spheres represent the N impurity. (c) and (d) show the spin density distribution in the planes passing through the graphene sheets in the corresponding models at left side. The red contours represent positive spin density and the green and blue ones represent higher and lower negative spin density. Contour plots of the charge density in the $(\bar{2}110)$ slices of $Gr^{AC}N^A$ and $Gr^{AC}N^C$ supercells are shown in (e) and (f) respectively. The slices become the diagonal lines cutting through the N atoms when the supercells are viewed from top as (a) and (b). (g) and (h) show spin-resolved PDOS of the N impurity and the neighboring C and Co in $Gr^{AC}N^{A}$ and $Gr^{AC}N^{C}$ respectively. Superscript of the notation in each pane denotes the site occupancy. Subscripts (1, 2 and 3) of the C notations denote the first, second and third nearest neighbors.